\documentclass[twocolumn,prl,superscriptaddress]{revtex4-1}
\usepackage{amsmath,amssymb,mathrsfs}
\usepackage{natbib}
\usepackage{subfigure}
\usepackage{tabularx}
\usepackage{epsfig}
\usepackage{longtable}
\usepackage{amsfonts}
\usepackage{rotating}
\usepackage{subfigure}
\usepackage{amsmath}
\usepackage{comment}
\usepackage{bbold} 
\usepackage{braket}

\def\be{\begin{equation}}
\def\ee{\end{equation}}
\def\bea{\begin{eqnarray}}
\def\eea{\end{eqnarray}}
\def\nn{\nonumber}

\usepackage{wordlike}

\def\ncsection#1{{\par{\vskip 5pt}\noindent\Large #1}}

\usepackage[unicode=true,bookmarks=true,bookmarksnumbered=false,bookmarksopen=false,breaklinks=false,pdfborder={0 0 1},backref=false,colorlinks=true]{hyperref}

\hypersetup{linkcolor=magenta,urlcolor=blue,citecolor=blue,pdfstartview={FitH},hyperfootnotes=false,unicode=true}

\begin{document}

\title{\bf Precise Programmable Quantum Simulations with Optical Lattices}
\author{Xingze Qiu}
\affiliation{State Key Laboratory of Surface Physics, Institute of Nanoelectronics and Quantum Computing, and Department of Physics, Fudan University, Shanghai, 200433, China}
\author{Jie Zou}
\affiliation{State Key Laboratory of Surface Physics, Institute of Nanoelectronics and Quantum Computing, and Department of Physics, Fudan University, Shanghai, 200433, China}
\author{Xiaodong Qi} 
\affiliation{ICIQ Research, Hangzhou, China} 
\author{Xiaopeng Li}
\email{xiaopeng\_li@fudan.edu.cn}
\affiliation{State Key Laboratory of Surface Physics, Institute of Nanoelectronics and Quantum Computing, and Department of Physics, Fudan University, Shanghai, 200433, China}
\affiliation{Shanghai Qi Zhi Institute, Shanghai, 200030, China}
\date{\today}

\begin{abstract}
We present an efficient approach to precisely simulate tight binding models with optical lattices, based on programmable digital-micromirror-device (DMD) techniques. 
Our approach consists of a subroutine of Wegner-flow enabled precise extraction of a tight-binding model for a given optical potential, 
and a reverse engineering step of adjusting the potential for a targeting model, for both of which we develop classical algorithms to achieve  high precision and high efficiency. 
With renormalization of Wannier functions and high band effects systematically calibrated in our protocol, we show the tight-binding models with programmable onsite energies and tunnelings can be precisely simulated with optical lattices integrated with the DMD techniques.  
With numerical simulation, we demonstrate that our approach would facilitate quantum simulation of localization physics with unprecedented programmability and 
atom-based boson sampling for illustration of quantum computational advantage. We expect this approach would  pave a way towards large-scale and precise programmable quantum simulations based on optical lattices.
\end{abstract}

\maketitle

Quantum simulation and quantum computing have been attracting tremendous attention in recent years. Among the rapidly advancing quantum hardwares~\cite{altman2019quantum}, 
cold atoms provide a unique quantum simulation platform for their controllability and scalability~\cite{bloch2012quantum, georgescu2014quantum, gross2017quantum, 2018_Bloch_quantum}. 
In the last two decades, cold atom based quantum simulations  have achieved fantastic progress not only along the line of conceptually novel physics such as artificial gauge fields~\cite{2011_Dalibard_RMP,zhai2015degenerate,Zhang_2018}, and topological matters~\cite{2019_Cooper_RMP}, but also along the line of simulating computationally difficult problems  such as BEC-BCS crossover~\cite{2004_Jin_PRL}, High-Tc physics~\cite{2002_Zoller_PRL,2013_Greif_Uehlinger_Science,mazurenko2017cold,2015_Hart_Duarte_Nature,2019_Bakr_BadMetal}, and non-equilibrium dynamics~\cite{2012_Bloch_Relaxation}, where its exceptional quantum advantage has been demonstrated. 

In quantum simulations aiming for demonstration of novel physical concepts, it is not crucial to precisely calibrate the system. However, in order to use quantum simulations to solve computationally difficult problems, it is required to make the simulation precise---for example in the study of quantum criticality and in  solving spin-glass problems, the physical properties of interest are sensitive to Hamiltonian parameters. And in quantum simulations of many-body localization using an incommensurate optical lattice, it has been found that  calibration problems cause qualitative disagreement~\cite{schreiber2015observation,li2017mobility,luschen2018single,kohlert2019observation} with the targeting Aubry-Andre (AA) model~\cite{harper1955single, aubry1980analyticity}.  This issue also arises generically in using speckle-pattern induced disorder optical potentials to simulate localization physics~\cite{damski2003atomic, gavish2005matter, schulte2005routes, billy2008direct, white2009strongly, sanchez2010disordered, pasienski2010disordered, kondov2011three, jendrzejewski2012three, semeghini2015measurement, smith2016many, choi2016exploring}, as the onsite energies and tunnelings are not programmable, let alone the simulation precision.  

Here we consider integration of  the recently developed DMD techniques in controlling optical potentials~\cite{ha2015roton, gauthier2016direct, 2016_Weiss_Science, mazurenko2017cold, browaeys2020many} to optical lattices, and calibrate the platform towards precise programmable  quantum simulations. 
{We develop an efficient algorithm,} 
which can systematically construct an inhomogeneous optical potential to precisely simulate a given tight binding lattice model, i.e., both the onsite energies and the tunnelings are made precisely programmable. 
{Its efficiency relies on the physical locality.}  
For benchmarking, we provide detailed numerical results for AA and Anderson localization (AL) models, where we show our approach has adequate programmability and systematically eliminates calibration errors. 
We show that our approach can also be used to implement atom-based quantum sampling algorithms such as boson sampling~\cite{aaronson2013computational,gard2015introduction} and determinantal point process~\cite{2012_Taskar_arXiv,li2019quantum}, having promising applications to  quantum machine learning. 
Our protocol provides precise programmability to the quantum platform of optical lattice,  which is intrinsically demanded for quantum simulations aiming for computationally difficult problems. 

\medskip
\ncsection{Results} 

{\bf Theory setup.} 
For atoms confined in an optical potential, the Hamiltonian description is 
\be 
H =-\frac{\hbar^2}{2m}\frac{d^2}{dx^2}+V_p (x)  + V_D (x) . 
\label{eq:Hamiltonian}
\ee 
Here we have separated the optical potential into a primary part $V_p (x) = \frac{V_p}{2}\cos(2kx)$ created by standard counter propagating laser beams and an additional potential $V_D(x)$ created by DMD~\cite{ha2015roton, gauthier2016direct, 2016_Weiss_Science, mazurenko2017cold, browaeys2020many} or sub-wavelength potential~\cite{Yi_2008} techniques. 
The primary part has lattice translation symmetry with the lattice spacing determined by the forming laser wavelength. 
Hereafter, we use the lattice constant $a=\pi/k$ as the length unit and the photon recoil energy of the lattice $E_R=\hbar^2k^2/2m$ as the energy unit. 
The added potential $V_D(x)$ in general has no homogeneity, and with the present technology it is typically much weaker than the primary lattice. 
A targeting tight-binding Hamiltonian matrix for the continuous system to simulate is referred to as ${\cal H}^\star$, 
which contains onsite energies $\epsilon_i$ and tunnelings ${J_{\langle ii'\rangle}}$, with ${i,i'}$ labeling lattice sites determined by the primary optical potential. 
In the following, we describe our numerical method to reverse engineer $V_D(x)$ and $V_p(x) $ that makes the precise  tight-binding model description  of $H$ in Eq.~\eqref{eq:Hamiltonian}   our target, ${\cal H}^\star$.

Firstly, we describe our method for efficient extraction of a tight-binding model of the continuous Hamiltonian $H$. 
Without the inhomogeneous potential $V_D(x)$, the precise tight binding model of the system can be efficiently constructed by introducing Bloch modes, because different modes with different lattice momenta are decoupled due to lattice translation symmetry. 
In the Wannier function basis, the Hamiltonian takes a block diagonalized form with the decoupled blocks corresponding to different bands~\cite{2016_Li_RPP}. 
In the presence of an inhomogeneous potential $V_D(x)$, the lattice translation symmetry becomes absent, and the Wannier states are coupled within each band and also across different bands. 
We propose to use Wegner flow~\cite{wegner1994flow, kehrein2006flow} to decouple different bands, which then produces a precise tight-binding model. 
We denote the Hamiltonian matrix in the Wannier function basis as ${\cal H}_{mi;m'i'}$, with 
{{$m,m'$}}
 labeling different bands running from zero (lowest band) to a high-band cutoff $M_c$, and 
{{$i,i'$}}
  the Wannier function localized centers (or equivalently the lattice sites of the primary lattice). 
The band decoupling procedure follows a flow equation, 
\begin{equation}
\frac{d{\cal H}(l)}{dl}=[\eta(l), {\cal H} (l)],
\end{equation}
that generates a continuous unitary transformation ${\cal H }(l) = U(l) {\cal H} (0) U^\dag (l) $. Here $\eta(l)$ is an anti-Hermitian matrix, 
$\frac{dU(l)}{dl}U^\dag(l)$, which we choose to be $\eta(l)=[{ G}, {\cal H}(l)]$, with 
${G}_{mi;m'i'} =\delta_{ii'}  {\left[2\delta_{mm'}-\delta_{m,0} \delta_{m',0} \right] }$. 
Following the flow from $l =0 $ to $+\infty$, 
${\cal H} (l)$ converges to a matrix that commutes with $G$ because 
\bea 
& \text{Tr}[{\cal H}(l)-G]^2 \ge 0 , &\\
& \frac{d}{dl}\text{Tr}[{\cal H}(l)-G]^2 
 =-2\text{Tr}[\eta^\dag (l) \eta (l) ]\le 0 &.  \nn 
\eea 
This means the coupling between the $m=0$ block of the matrix ${\cal H}$ and other blocks monotonically converges to $0$.  
A more thorough analysis shows an exponential convergence with a convergence speed inversely proportional to the band gap (see Methods).
This means our approach is applicable as long as the inhomogeneous potential $V_D(x)$ is not too strong to close the band gap. 
The finite-depth flow equation generates a local unitary that defines a precise tight-binding model as the converged  $m=0$ Hamiltonian block.

Secondly, we develop a numerical optimization method to adjust the potential $V_D (x)$ to minimize the difference between ${\cal H}_{\rm eff}$ and ${\cal H}^\star$. 
We choose a Frobenius-norm based cost function  $f = f_0 + \lambda_1f_1$, where 
${f}_0$ and ${f}_1$ are Frobenius norms for the difference in  the onsite energies and tunnelings, respectively, 
and a hyper-parameter $\lambda_1$ is introduced to afford extra weight to the tunneling for better  optimization-performance. 
In our numerics, we parameterize 
\be 
V_D(x) = \sum^{2L-1}_{n=0}\frac{\widetilde{V}_n}{2}\cos\left(2\frac{n}{L}kx+\widetilde{\phi}_n\right),
\ee
where $L$ is the number of periods of the primary lattice, 
and $\widetilde{V}_n$, $\widetilde{\phi}_n$ are variational parameters.
We start from a random initialization, obtain ${\cal H}_{\rm eff}$ through Wegner flow, and then update the optical potential through a gradient descent method. 
This procedure is iterated until the cost function is below a threshold of our request.

Furthermore, our method is highly efficient by making use of locality. 
Considering a system with large system size,
instead of performing the Wegner-flow for the full problem which then has a computation complexity of $O(L^3)$, we split the system into small pieces, with an individual length $L_p$. The adjacent pieces have about one third of their length overlapped with each other. We optimize the optical potential to reproduce the precise tight-binding model piece-by-piece, and then glue them together. This is sensible because of the locality in the problem---the onsite energy at one site and the tunnelings between two sites are both determined by their neighboring potential, following the finite-depth Wegner flow.
Note that one problem arises that the potential may not be smooth in the overlapping regions, as the obtained potential could be inconsistent in optimizing the two adjacent pieces. 
To solve this  problem,  we add $\lambda_2 {f}_2$ to the cost function, where ${f}_2$ is the Frobenius norm of the difference of the 
 potential in the overlapping region obtained in the optimization of its belonging two pieces (see Methods). The piece-by-piece procedure is swept back-and-forth for convergence, analogous to the optimization in the standard  density-matrix-renormalization group calculation~\cite{white1992density}. In the sweeping process, we find a monotonic decrease in the difference between ${\cal H}_{\rm eff}$ and ${\cal H}^\star$ in the whole system, and that the converged optical potential is smooth.  The computation complexity scaling is thus reduced to $O(L)$.

\begin{figure}
\includegraphics[width=\linewidth]{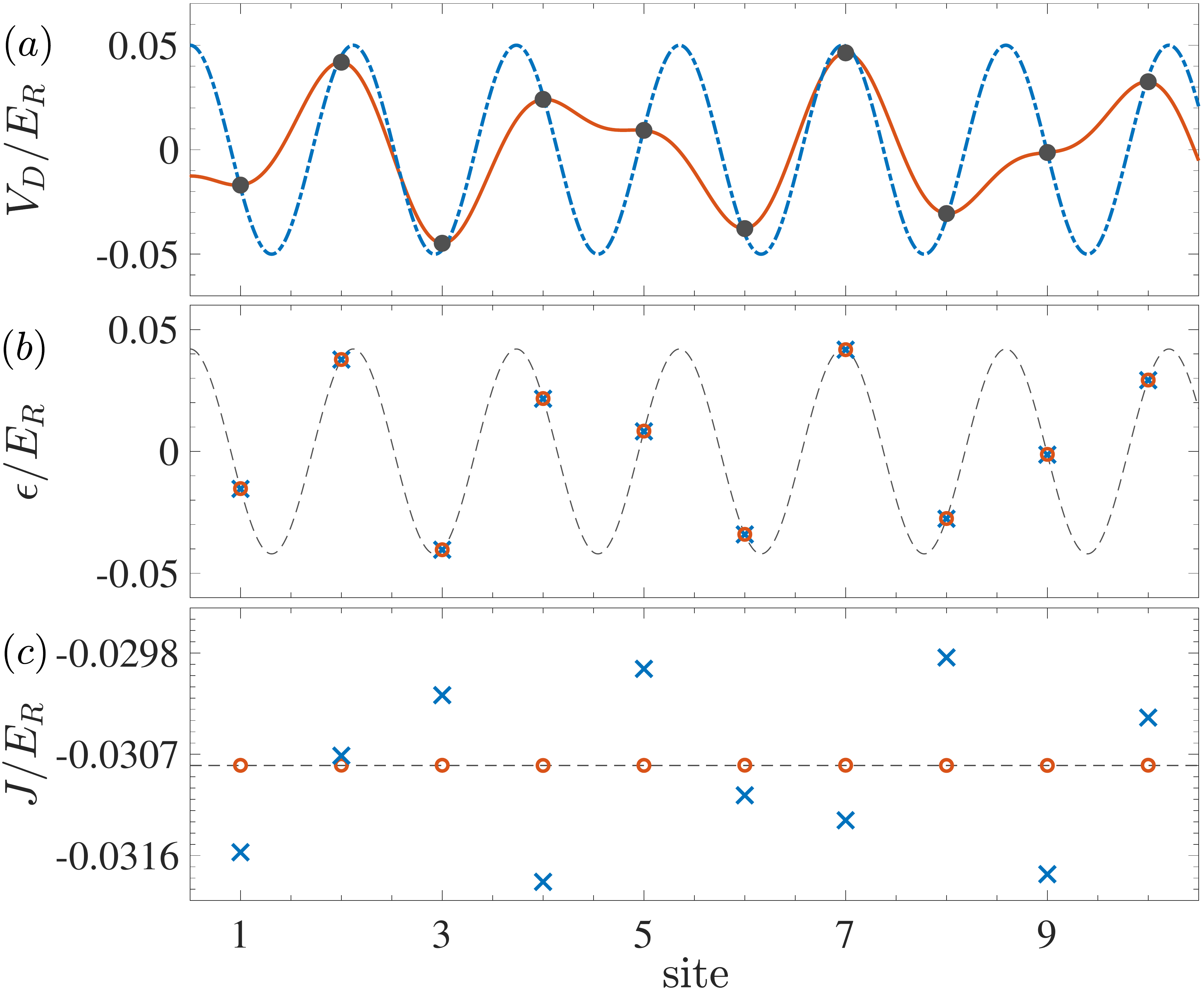}
\caption{
Precise quantum simulation of AA model.
(a), The optical potential $V_{D, precise}$ (red solid line) and $V_{D, err}$ (blue dash-dotted line).  
The potential $V_{D, precise}$ possesses a vanishing derivative at the individual sites (black dots), whereas $V_{D, err}$ does not. (b), The onsite energies produced by $V_{D, precise}$ (red circles) and $V_{D, err}$ (blue crosses). 
The dashed line is the desired sinusoidal form of the site-dependent onsite energies in $H^\star_{AA}$. 
(c), The tunnelings produced by $V_{D, precise}$ (red circles) and $V_{D, err}$ (blue crosses). 
The dashed line marks  the desired site-independent tunnelings in $H^\star_{AA}$. 
Here, we choose the high-band cutoff $M_c = 2$,  the hyper-parameter $\lambda_1=1$, the system size $L=55$, and periodic boundary condition.
}
\label{fig:AAmodel}
\end{figure}

{\bf Application to quantum simulation of AA model.}
In the study of quantum localization physics, AA model has been investigated extensively in both theory and experiment~\cite{grempel1982localization, roati2008anderson, sanchez2010disordered, deissler2010delocalization, biddle2010predicted, schreiber2015observation, li2015many, modak2015many, li2017mobility, luschen2018single, kohlert2019observation, lukin2019probing}. 
Its Hamiltonian reads as 
\bea 
\textstyle H^\star_{AA} =&-&
\textstyle J_{AA}\sum_i\left(c^\dag_{i+1}c_i+h.c.\right)\nn \\
&+&
\textstyle \frac{\epsilon_{AA}}{2}\sum_i\cos(2\pi\alpha i+\phi)c^\dag_ic_i,
\eea
where $c^\dag_i$ ($c_i$) denotes the creation (annihilation) operator on a lattice site $i$, $\alpha$ is an irrational number, $J_{AA} $ is the site-independent tunneling, 
$\epsilon_{AA}$ describes the strength of the onsite energies, and $\phi$ is an arbitrary phase. Here, we choose $\alpha$ as the golden ratio $(\sqrt{5}-1)/2$, which is approximated by the Fibonacci sequence ($F_n$) as    $F_n/F_{n+1}$ in a finite-size calculation. 
Because of its energy independent duality defined by a Fourier transform, the model exhibits a phase transition from all wave-function localized to all extended, which makes it natural place to examine one-dimensional localization criticality. 

In the optical lattice experiment~\cite{schreiber2015observation}, the AA model Hamiltonian is achieved by using an incommensurate bichromatic potential, a primary lattice perturbed by a second weak incommensurate lattice with  $V_{D, err} (x) = V_1 \cos(2\alpha kx)/2$ following our notation in Eq.~\ref{eq:Hamiltonian}. However, its corresponding tight-binding model is not a precise AA model---there are corrections making  tunnelings inhomogeneous and generating higher-order harmonics, which generically breaks the central ingredient of duality of the AA model~\cite{2015_Ganeshan_PRL}. The effects of such corrections have been established both in theory~\cite{li2017mobility} and experiment~\cite{luschen2018single,kohlert2019observation}. This problem can be solved by using our precise quantum simulation method.

Through the optimization described above, we find that precise quantum simulation of AA model is achieved by choosing a potential 
\be 
V_{D, precise} (x) = 
 \frac{\widetilde{V}_1}{2}\cos\left(2\frac{F_n}{F_{n+1}}kx\right) 
+\frac{\widetilde{V}_2}{2}\cos\left(2\frac{F_{n-1}}{F_{n+1}}kx\right),
\ee 
with appropriate coefficients $\widetilde{V}_{1,2}$. 
As an example, we consider a specific model $H^\star_{AA}$ with parameters $J_{AA} = -0.0308E_R$, $\epsilon_{AA} = 0.0841E_R$, $\alpha \approx F_n/F_{n+1}$  ($F_n = 34$ and $F_{n+1} = 55$), and $\phi=-\pi\alpha$. 
This target model is reached by choosing $V_p = 8E_R$, $V_1 = 0.1E_R$, $\widetilde{V}_1 = 0.0341E_R$, and $\widetilde{V}_2 = -0.0592E_R$. 
In Fig.~\ref{fig:AAmodel}(a), we show the optical potentials corresponding to $V_{D, err}$ and $V_{D, precise}$ for comparison.
We find that the resultant onsite energies are approximately the same (Fig.~\ref{fig:AAmodel}(b)), yet with the potential $V_{D, precise}$ giving a more precise solution.
More drastically, the tunnelings out of our potential with $V_{D, precise} (x)$ are precisely homogeneous, with a relative inhomogeneity below $1E-4$ (Fig.~\ref{fig:AAmodel}(c)). This cannot be achieved with the potential of $V_{D, err} (x)$.

We also emphasize here that our constructed potential $V_{D, precise} (x)$ possesses a vanishing derivative at the individual sites, as exhibited in Fig.~\ref{fig:AAmodel}(a). This is crucial to  experiments as a potential with finite derivative at the position of atoms would make the system more susceptible to shaking-induced heating processes ~\cite{lukin2019probing}.

\begin{figure*}
\includegraphics[width=\linewidth]{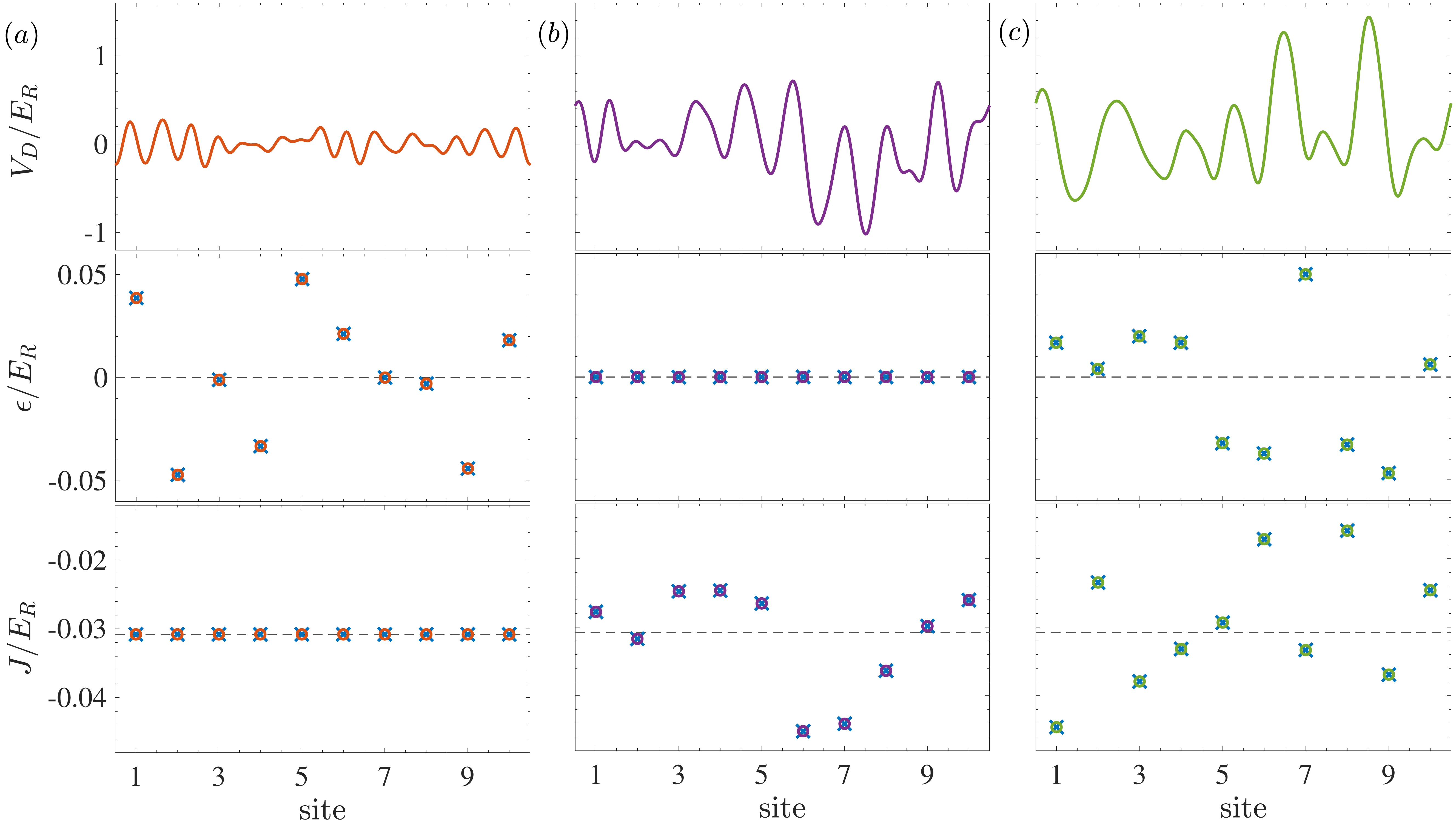}
\caption{
Precise quantum simulation of three cases of AL models. 
(a), The random onsite model with $h_i\in[-0.05E_R,0.05E_R]$ and $t_i= -0.0308E_R$ being homogeneous. (b),  The random hopping model with $h_i=0$ homogenous, $t_i\in[J_{AL}-|J_{AL}|/2,J_{AL}+|J_{AL}|/2]$, and $J_{AL}= -0.0308E_R$. (c),   Both onsite energy and tunneling being random with $h_i\in[-0.05E_R,0.05E_R]$, $t_i\in[J_{AL}-|J_{AL}|/2,J_{AL}+|J_{AL}|/2]$, and $J_{AL}= -0.0308E_R$. 
The first row shows the reverse-engineered optical potentials $V_D(x)$, the middle row and the last row shows the onsite energies and the tunnelings. 
Circles and crosses indicate the values in the tight-binding models extracted from the continuous Hamiltonian in Eq.~\eqref{eq:Hamiltonian} and the targeting tight binding model,  respectively.
Here, we choose the high-band cutoff $M_c = 2$,  the hyper-parameter $\lambda_1=100$, the system size $L=10$, the depth of the primary lattice $V_p=8E_R$, and periodic boundary condition.
}
\label{fig:Anderson}
\end{figure*}

\begin{figure}
\includegraphics[width=.95\linewidth]{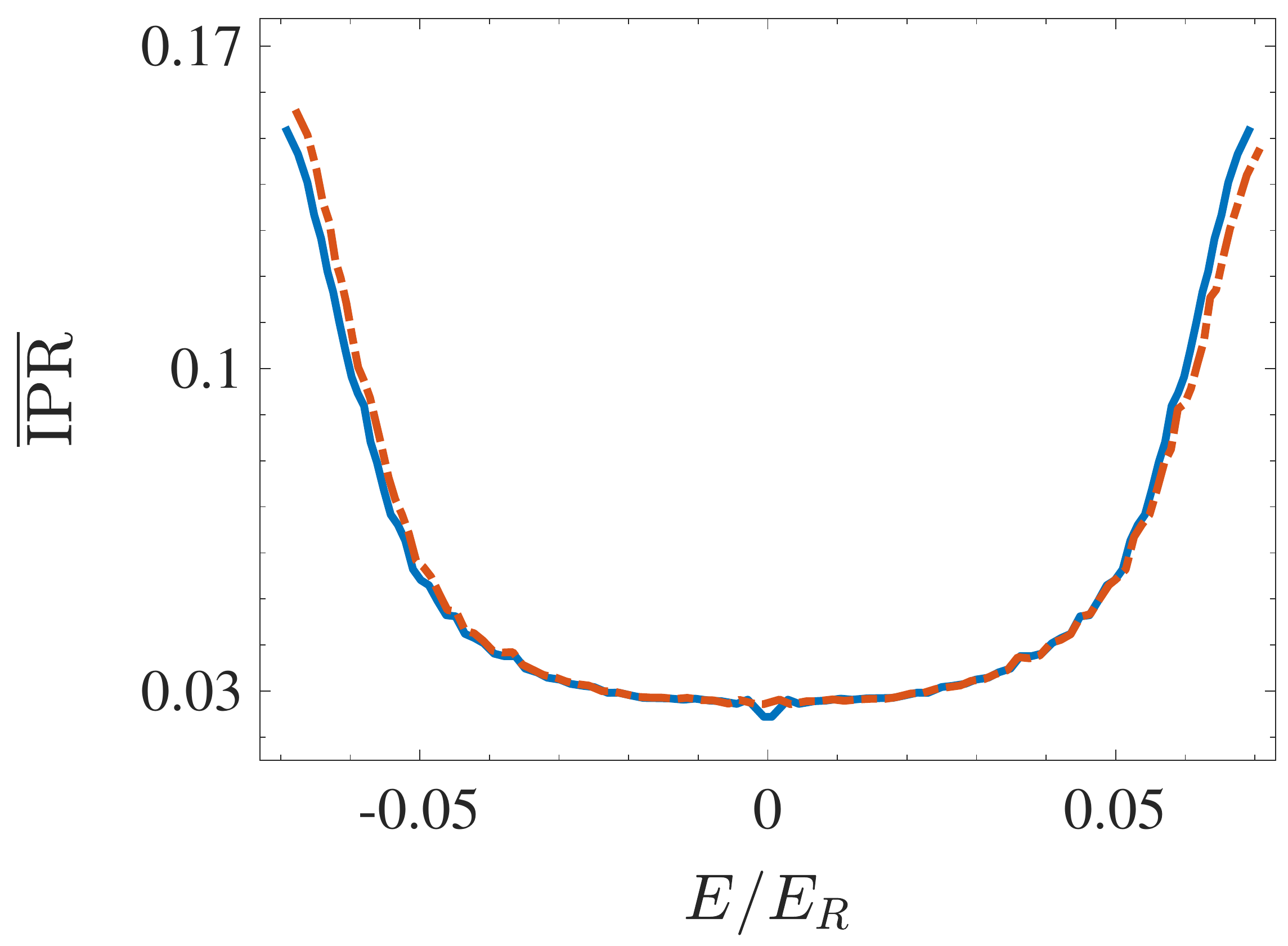}
\caption{Averaged IPR for the random hopping model (Eq.~\eqref{eq:ALmodels}) by sampling 2000 disorder configurations. Here we set $J_{AL}=-0.0308E_R$ and $\widetilde{J}_{AL} =2|J_{AL} |/3$, the system size $L = 100$. The blue solid, and red dash-dotted lines, correspond to the results obtained from diagonalizing the continuous Hamiltonian in Eq.~\eqref{eq:Hamiltonian} and the tight binding random-hopping model, respectively.  
Here, we choose the high-band cutoff $M_c = 2$, the hyper-parameter $\lambda_1=100$, the system size $L=100$, the depth of the primary lattice $V_p=8E_R$, and periodic boundary condition. The whole system is split into a number of pieces with $L_p=10$, and the adjacent pieces overlap with each other over $4$ sites. 
We choose the hyper-parameter $\lambda_2=0.5$.
}
\label{fig:hopping}
\end{figure}

{\bf Anderson localization with programmable disorder potential.}
To further demonstrate the precise programmability enabled by our method, we also carry out an application to quantum simulation of Anderson localization models whose previous experimental realization by speckle pattern lacks programmability~\cite{damski2003atomic, gavish2005matter, schulte2005routes, billy2008direct, white2009strongly, sanchez2010disordered, pasienski2010disordered, kondov2011three, jendrzejewski2012three, semeghini2015measurement, smith2016many, choi2016exploring}.
The Hamiltonians of 1D AL models are given as
\begin{equation}
H^\star_{AL}=\sum_ih_ic^\dag_ic_i+\sum_i\left(t_ic^\dag_{i+1}c_i+h.c.\right). 
\label{eq:ALmodels}
\end{equation}
We consider three different cases: (a) random onsite model with $h_i\in[-\widetilde{\epsilon}_{AL} /2, \widetilde{\epsilon}_{AL}/2]$ and $t_i=J_{AL}$ being homogeneous, (b) random hopping model with $h_i=\epsilon_{AL}$ homogenous and $t_i\in[J_{AL}-\widetilde{J}_{AL}/2,J_{AL}+\widetilde{J}_{AL}/2]$, and (c) 
both onsite energies and tunnelings being random with $h_i\in[ -\widetilde{\epsilon}_{AL} /2, \widetilde{\epsilon}_{AL}/2]$ and $t_i\in[J_{AL} -\widetilde{J}_{AL}/2,J_{AL}+\widetilde{J}_{AL}/2]$. The random onsite energies and tunnelings are drawn according to a uniform distribution. 
In Fig.~\ref{fig:Anderson}, we show all the three different cases of AL model can be precisely achieved with our optimization method.  The absolute errors in the tight-binding model compared to the target one  is made  smaller than $1E-5$, which demonstrates the precise programmability of our scheme.

One immediate application of the programmable quantum simulation of Anderson localization is to study the anomalous localization in the random hopping model. Unlike the random onsite model, where all states are localized in one dimension,  the random hopping model has delocalized states at band center~\cite{eggarter1978singular, balents1997delocalization}. 
But it is extremely difficult to perform quantum simulation of  this pure random hopping model with the speckle-pattern approach lacking programmability, since the unavoidable  inhomogeneity in the  onsite energy will make all states localized. We randomly generate $2000$ disorder samples for the hopping, and compute the corresponding potential $V_D (x)$ using our optimization method. The averaged inverse participation ratio (IPR) which diagnoses localization to delocalization transition~\cite{vadim} is calculated,  with the results shown in Fig.~\ref{fig:hopping}. 
We find quantitative agreement of results obtained for the continuous potential with the targeting tight-binding model. 
The discrepancy can be further improved by increasing the lattice depth or allocating more numerical resources.

{\bf Implementation of boson sampling and determinantal point process.}
Boson sampling is a promising candidate to demonstrate quantum  computational advantage for its established exponential complexity on a classical computer \cite{aaronson2013computational,gard2015introduction,clifford2018classical}. Its experimental implementation  has been achieved in linear photonic~\cite{Flamini_2018},  trapped ion \cite{shen2014scalable}, and quantum-dot devices \cite{he2017time}. 
Here we show that  boson sampling could also be implemented with bosonic atoms confined in an optical lattice using our developed precise programmability. 
One advantage of atomic realization is that one can replace bosonic atoms by their fermionic isotopes, which then performs quantum sampling for determinantal point process~\cite{2012_Taskar_arXiv}. This then provides one way to verify the quantum advantageous boson sampling because the simulation of determinantal point process is efficient on a classical computer~\cite{2012_Taskar_arXiv,li2019quantum}. 


Here, we consider a standard boson sampling problem with $m$ input modes and $n$ identical bosons, where the $n$ bosons are one-to-one injected into the first $n$ modes as the input state, and 
{then let evolve under an} 
$m \times m$ Haar-random unitary $U$. In the dilute limit ($n \ll m$), where each output mode contains at most one particle, the probability of a specific output Fock-state configuration $S$ is $p(S)=|\text{per}(U_S)|^2$, with $\text{per}$ meaning the permanent, and $U_S$ a submatrix of $U$ selected according to the input and output configurations~\cite{aaronson2013computational}.

\begin{figure}
\includegraphics[width=\linewidth]{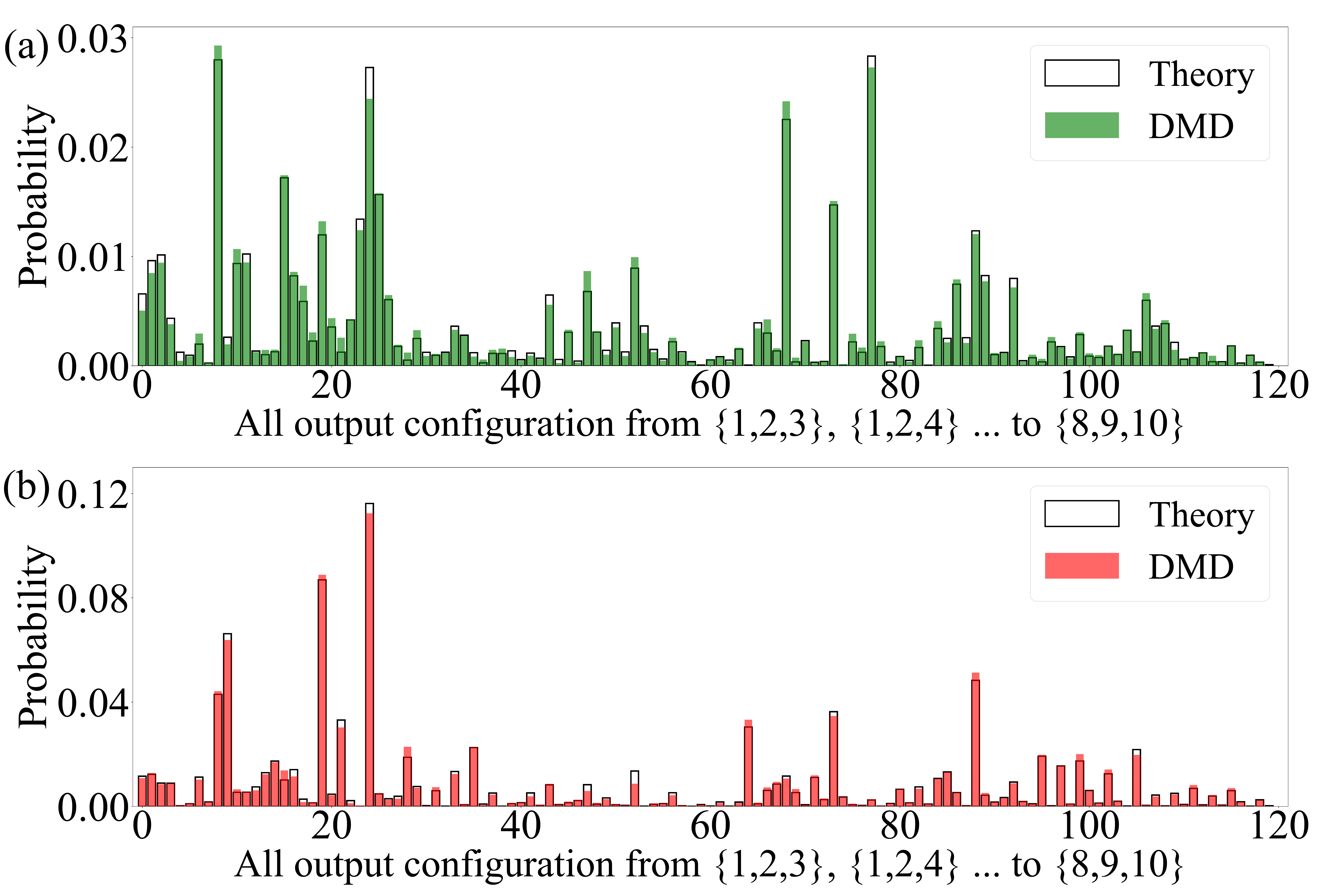}
\caption{
Comparison between theory and simulated experimental realization for boson sampling and determinantal point process. 
Here we choose mode number $m = 10$ and particle number $n = 3$. 
 The solid bars indicate the results of the simulated experimental realization based on DMD enabled programmability. 
Empty bars indicate the results from the precise targeting theory model. 
The similarities (distances) between them are (a) $G=0.997$ $(D=0.0525)$ and (b) $G=0.998$ $(D=0.0412)$. Here, we only show no-collision output combinations in the bosonic case, and we set $J_x=-0.01E_R$, $\epsilon_z=0.2E_R$.
}
\label{fig:sampling}
\end{figure}

To experimentally realize the Haar-random unitary $U$ with an optical lattice, we 
{adapt} 
the decomposition in Ref.~\onlinecite{michael1994experimental}, where the random unitary is constructed by multiplication of a series of building blocks of two-mode unitary operations. For the optical lattice implementation, we develop a different construction from photonic realization~\cite{wang2017high} (see Methods).  We choose the two-mode building blocks as 
\be
T^{(p,q)}=\exp\left(ih^{(p,q)}_z\sigma^{(p,q)}_z\right)\exp\left(ih^{(p,q)}_x\sigma^{(p,q)}_x\right). 
\ee
Here we have  the Pauli matrices $\sigma^{(p,q)}_x=\ket{q}\bra{p}+\ket{p}\bra{q}$ and $\sigma^{(p,q)}_z=\ket{q}\bra{q}-\ket{p}\bra{p}$, $p,\, q\in\{1,2,\cdots,m\}$,  the quantum states $|p\rangle$ and $|q\rangle$ represent the Wannier functions in the optical lattice, 
and $h^{(p,q)}_{x,z}$ are parameters determined by $U$. 
With our optimization method, we can obtain the Hamiltonians $H^{(q+1,q)}_z=\epsilon_z\sigma^{(q+1,q)}_z$ and $H^{(q+1,q)}_x=J_x\sigma^{(q+1,q)}_x$, and hence the unitary $\left(T^{(q+1,q)}\right)^\dag$ can be achieved through the time evolution operator
$
\exp\left(-i\tau^{(q+1,q)}_xH^{(q+1,q)}_x\right)\exp\left(-i\tau ^{(q+1,q)}_zH^{(q+1,q)}_z\right),
$
with the evolution time $\tau ^{(q+1,q)}_x=h^{(q+1,q)}_x/J_x$ and $\tau^{(q+1,q)}_z=h^{(q+1,q)}_z/\epsilon_z$, which are positive with proper construction (see Methods). 
For $p-q>1$, a more involved construction is required, which is provided in Methods. The building blocks of $T^{(p,q)}$ ultimately realize any random unitary. 
As concrete examples, we consider $^{87}\text{Rb}$ atoms confined in a lattice formed by a laser with wavelength $1064\, \text{nm}$. 
For a mode number $m=10$ and number of atoms $n=3$, the total evolution time is estimated to be $1.2$ seconds, which is accessible with the current lifetime of cold atoms.  
Denoting the probabilities corresponding to the theory and the simulated DMD-based experimental realization as $p_1(S)$ and $p_2(S)$, respectively,  the sampling precision is characterized by  a measure of similarity $G=\sum_S\sqrt{p_1(S)p_2(S)}$, and a measure of distance $D=(1/2)\sum_S|p_1(S)-p_2(S)|$. The numerical results are shown in Fig.~\ref{fig:sampling} (a). We find quantitative agreement between the simulated experimental realization and the theory prediction, which implies the precision achieved with our scheme is adequate to perform boson sampling experiments. 

We also study the case with fermionic atoms, which then realize the determinantal point process~\cite{2012_Taskar_arXiv}. 
The results are shown in Fig.~\ref{fig:sampling}(b), where we also find the quantitative agreement between the simulated experimental realization and the theory prediction. It is worth noting here that even when the classical simulation of boson sampling is unavailable for a large particle number, the experiment with fermions allows one way to verify the quantum device as the determinantal point process is efficiently simulatable on a classical computer~\cite{2012_Taskar_arXiv,li2019quantum}.

\ncsection{Discussion} 

We have proposed a scheme for precisely simulating lattice models with optical lattices, whose potentials can be  manipulated through the high-resolution DMD techniques. 
We have developed a Wegner-flow  method to extract the precise tight-binding model of a continuous potential,
and a scalable optimization method for the reverse engineering of the optical potential whose tight binding model precisely matches a targeting model. 
The performance is demonstrated with concrete examples of AA and Anderson models, and quantum sampling problems. 
Our approach implies optical lattices can be upgraded towards high-precision programmable quantum simulations by integrating with DMD techniques.

The precise programmable quantum simulation enabled by our scheme make the optical lattice rather flexible. For disorder physics, having programmable disorder allows for more systematic study of the localization transition, especially for cases where the rare disorder Griffith effects are important for example in understanding disordered Weyl semimetals~\cite{pixley2016rare}, and many-body localization mobility edge~\cite{2006_Basko_MBL, 2016_Muller_PRB}. 
Our proposing setup also paves a way to building a programmable quantum annealer with optical lattices. Considering spinor atoms in a deep lattice with strong interaction,  programmable tunnelings imply programmable spin-exchange. 

\medskip 
\ncsection{Methods}

{\bf Exponential convergence of Wegner flow.} 
As the efficiency of our method relies on the convergence behavior of Wegner flow, in this section we prove that the convergence is exponential, and that the convergence speed is inversely proportional to the band gap---it is  lower bounded by a value inversely proportional to the band gap to be more precise. 

Note that we use Wegner flow to decouple the lowest band from the rest. The flow converges when the coupling between the lowest and excited bands vanish. To analyze such  couplings, we rewrite the Hamiltonian matrix in terms of the lowest and excited band blocks and their couplings as 
\bea 
&&{\cal H} _{mi; m'i'} 
= {\cal D} ^{(0)}_{ii'} \delta_{m0} \delta_{m'0} + {\cal D} ^{(1)}_{mi; m'i'} (1-\delta_{m0}) (1-\delta_{m'0}) \nn \\ 
&&+ \delta_{m0} {\cal C} _{i; m'i'} (1-\delta _{m'0}) +  (1- \delta_{m0} ){\cal C} ^*_{i'; mi} \delta_{m'0}.   
\eea 
Following our constructed Wegner flow, we have the flow equation for the coupling matrix ${\cal C}$ as 
\be 
\frac{d {\cal C}} {dl} = {\cal D} ^{(0)} {\cal C} - {\cal C} {\cal D} ^{(1)}. 
\ee 

The overall strength of these couplings in ${\cal C}$  are quantified by the trace ${\rm Tr} [{\cal C} ^\dag {\cal C} ]$, whose $l$-dependence obeys 
\bea 
  \frac{d}{dl} {\rm Tr}  [{\cal C}^{\dag} {\cal C} ]  
&&= 2 {\rm Tr}  [ {\cal C} {\cal C}^{\dag} {\cal D}^{(0)} - {\cal C} ^{\dag } {\cal C} {\cal D}^{(1)}] \nn \\ 
&&< - 2 \left[ d_{\rm min}^{(1)} - d_{\rm max} ^{(0)}  \right] {\rm Tr} [{\cal C} ^\dag {\cal C} ], 
\eea 
with $d_{\rm min} ^{(1) } $ the minimal eigenvalue of the Hermitian matrix ${\cal D}^{(1)}$ and $d_{\rm max}^{(0)} $ the  maximal eigenvalue of ${\cal D} ^{(0)}$. 
We then obtain a  bound on the trace as 
\be 
{\rm Tr}  [{\cal C} ^\dag {\cal C}  ]_{l_0+\Delta l } < {\rm Tr}   [ {\cal C} ^\dag {\cal C} ]_{l_0} e^{-2[ d_{\rm min}^{(1)} - d_{\rm max} ^{(0)}] \Delta l} . 
\ee 
Having a finite gap between the lowest and the first excited bands, we have $d^{(1)}_{\rm min} - d_{\rm max} ^{(0) } > 0 $.  The exponential convergence of the couplings between the lowest and excited bands is then assured. The convergence speed is larger than a value inversely proportional to the band gap.

\begin{figure*}
\includegraphics[width=\linewidth]{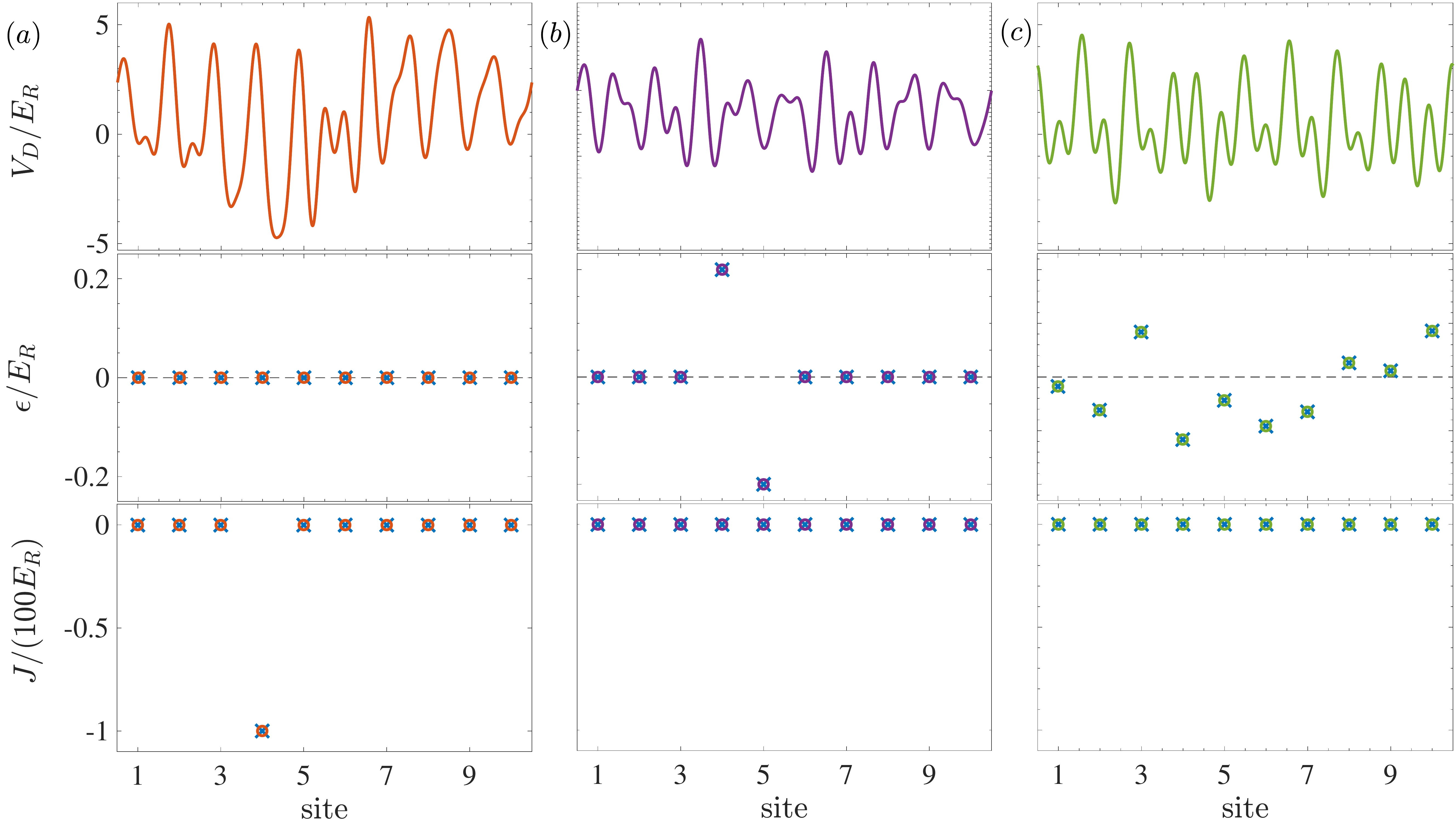}
\caption{
Precise quantum simulation of the Hamiltonians corresponding to different quantum gates. 
(a), $H^{(5,4)}_x = J_x\sigma^{(5,4)}_x$ with $J_x= -0.01E_R$. (b),  $H^{(5,4)}_z = \epsilon_z\sigma^{(5,4)}_z$ with $\epsilon_z= 0.2E_R$. (c), 
$H_d=\frac{i}{8\pi}\log(U_{\rm diag})$, with $U_{\rm diag}$ defined in Eq.~\eqref{eq:Um}, 
corresponding to the decomposition of the Haar-random unitary in Fig.~\ref{fig:sampling} (see the notation in Methods). 
The first row shows the reverse-engineered optical potentials $V_D(x)$, the middle row and the last row shows the onsite energies and the tunnelings. 
Circles and crosses indicate the values in the tight-binding models extracted from the continuous Hamiltonian and the targeting tight binding model,  respectively.
Here, we choose the high-band cutoff $M_c = 2$,  the hyper-parameter $\lambda_1=20$, and the depth of the primary lattice $V_p=20E_R$.
}
\label{fig:gates}
\end{figure*}

 {\bf The piece-by-piece optimization method.}
In this section, we provide the details of the piece-by-piece optimization method. Taking a system having $L$ number of periods---the period is defined according to the primary lattice, the starting points of the periods are labeled as $(X_0, X_1, X_2, \ldots, X_{L-1})$. 
We split the system into smaller pieces with a piece-size $L_p$. Two adjacent pieces have a finite overlap region with size $M_p$. The $i$-th piece contains the periods from $X_{i(L_p-M_p)}$ to $X_{i(L_p-M_p)+L_p}$.  Its overlap with the lefthand [righthand] side $(i-1)$-th [$(i+1)$-th] piece is from $X_{i(L_p-M_p)}$ to $X_{(i-1)(L_p-M_p)+L_p } $  [from $X_{(i+1) (L_p-M_p)} $ to $X_{i(L_p-M_p)+L_p}$].  
In optimizing the optical potential at $i$-th piece for the targeting tight-binding model in that local region, 
we introduce an additional cost function $\lambda_2 f_2$, 
with $\lambda_2$ a hyper-parameter, and 
\bea 
f_2 &=&  \sqrt{\int _{X_{i(L_p-M_p)}} ^{X_{(i-1)(L_p-M_p)+L_p } }   dx 
				 \left[V_{D, i} (x)-V_{D, i-1} (x)  \right] ^2} \nn \\ 
& +& \sqrt{\int_{X_{(i+1) (L_p-M_p)} }  ^{X_{i(L_p-M_p)+L_p}}   dx  
 				\left[V_{D, i} (x)-V_{D, i+1} (x)  \right] ^2}, 
\eea 
where $V_{D, i}(x)$ is the variational potential in optimizing the $i$-th piece. 
The $f_2$ cost function  is introduced to minimize the inconsistency of the potential in the overlap region with the neighboring  $(i-1)$-th and   $(i+1)$-th pieces. Since  the constraint on the consistency is not implemented strict, there will still be leftover inconsistency between $V_{D, i} (x)$ and $V_{D, i\pm1} (x)$ in a single run. To solve this problem, we perform a back-and-forth sweeping process---we first carry out optimization in a forward direction  from the leftmost piece to the rightmost,  and then in a backward direction from the rightmost to leftmost. This sweeping process is iterated for potential convergence. 
In our numerics, we find convergence with three to four  times of sweeping.  
We then glue all the pieces together and construct the global optical potential. It is confirmed that this procedure gives the correct potential whose tight binding model is the targeting model.

{\bf Decomposition of a Haar-random unitary with optical lattice accessible operations.}
Here we describe how to adapt the decomposition of the Haar-random unitary in Ref.~\onlinecite{michael1994experimental} 
to optical lattice implementation. An $m\times m$ Haar-random unitary $U(m)$ is decomposed into
\be
\textstyle U(m)=U_{\rm diag} \times \left[\left(\prod^2_{p=m}\prod^{1}_{q=p-1}T^{(p,q)}\right)\right]^{\dag}. 
\label{eq:Um} 
\ee
The order of matrix multiplication using $\prod$ is defined to be from left to right, for example $\prod_{i=3}^1 A_i$ means $A_3 A_2 A_1$ and $\prod_{i=1}^3 A_i$ means $A_1 A_2 A_3$. In the above equation, we have 
\be
\textstyle T^{(p,q)}=
\textstyle 
\exp\left(ih^{(p,q)}_z\sigma^{(p,q)}_z\right)\exp\left(ih^{(p,q)}_x\sigma^{(p,q)}_x\right),
\label{eq:Tpq}
\ee
where the Pauli operations are defined according to the Wannier basis quantum states $|q\rangle$ with $q$ the lattice site index---
$\sigma^{(p,q)}_z=\ket{q}\bra{q}-\ket{p}\bra{p}$, $\sigma^{(p,q)}_x=\ket{q}\bra{p}+\ket{p}\bra{q}$. 
To specify the matrix $T^{(p,q)}$, we introduce a matrix $\widetilde{U}^{(p,q)}$, whose elements $\widetilde{U}^{(p,q)}_{p,q}$ and $\widetilde{U}^{(p,q)}_{p,p}$ determine the parameters $h_{x,z} ^{(p,q)}$ as 
\bea
&&\textstyle h^{(p,q)}_x = -\arctan\left(\left|\frac{\widetilde{U}^{(p,q)}_{p,q}}{\widetilde{U}^{(p,q)}_{p,p}}\right|\right) \leqslant 0,  \nn \\ 
&& \textstyle h^{(p,q)}_z = \frac{1}{2}\left[\pi-\arg\left(\frac{i\widetilde{U}^{(p,q)}_{p,q}}{\widetilde{U}^{(p,q)}_{p,p}}\right)\right]\geqslant 0. 
\label{eq:hxz}
\eea
Here, $T^{(p,q)}$ and $\widetilde{U}^{(p,q)}$ are constructed in a sequential manner as $(p,q)$ goes through the sequence $\{(m,m-1),(m,m-2),\cdots,(m,1),(m-1,m-2),(m-1,m-3),\cdots,(2,1)\}$.
From $\widetilde{U}^{(m,m-1)}=U(m)$, we obtain $T^{(m,m-1)}$ through Eq.~\eqref{eq:hxz} and Eq.~\eqref{eq:Tpq}, and then we have $\widetilde{U}^{(m,m-2)}=\widetilde{U}^{(m,m-1)}T^{(m,m-1)}$. In general once $\widetilde{U}^{(p_1,q_1)}$ and $T^{(p_1,q_1)}$ are obtained, we have  $\widetilde{U}^{(p_2,q_2)}=\widetilde{U}^{(p_1,q_1)}T^{(p_1,q_1)}$ for $(p_2, q_2)$ next to $(p_1, q_1)$ in that sequence.  Following this sequence, all matrices are constructed. 
The additional matrix $U_{\rm diag} $ in Eq.~\eqref{eq:Um} is diagonal with the elements $(U_{\rm diag})_{n,n}=\left[\widetilde{U}^{(2,1)}T^{(2,1)}\right]_{n,n}$, $n=1,2,\cdots,m$.

From Eq.~\eqref{eq:Um}, we see that to realize the Haar-random unitary $U(m)$, the building block is the unitary $\left(T^{(p,q)}\right)^\dag$, which can be achieved through time evolution of the corresponding Hamiltonian, as specified latter. 
To engineer the non-local gate operation $\left(T^{(p,q)}\right)^\dag$
we perform the following transformation,
\be
\textstyle \sigma^{(p,q)}_{x,z} = \ket{q}\bra{p}+\ket{p}\bra{q} = \left(U^{(p,q)}\right)^\dag\sigma^{(q+1,q)}_{x,z}U^{(p,q)}, \nn 
\ee
where 
\be
U^{(p,q)}=\left\{\begin{array}{ccc}\text{Identity matrix}, & p=q+1, \\ \prod^{p-1}_{n=q+1} U^{(n+1,n)}, & p>q+1, \end{array}\right.
\nn
\ee
with
\bea
U^{(n+1,n)}=~&&\ket{n}\bra{n+1}-\ket{n+1}\bra{n} \nn \\
&&+ \sum_{k\in\{1,\cdots,m\}\backslash\{n,n+1\}}\ket{k}\bra{k}. 
\nn 
\eea
Hence we have
\be
T^{(p,q)} =  \left(U^{(p,q)}\right)^\dag\widetilde{T}^{(p,q)}U^{(p,q)}, 
\nn 
\ee
with 
\bea
\widetilde{T}^{(p,q)}= e^{i\tau^{(p,q)}_zH^{(q+1,q)}_z}e^{i\tau^{(p,q)}_x H^{(q+1,q)}_x}, 
\nn 
\eea
which corresponds to time evolution with tight binding Hamiltonians
\be
H^{(q+1,q)}_z = \epsilon_z\sigma^{(q+1,q)}_z,
~H^{(q+1,q)}_x = J_x\sigma^{(q+1,q)}_x. 
\nn 
\ee
Here $\epsilon_z>0$ and $J_x<0$ are constants, and the evolution time is $\tau^{(p,q)}_z=h^{(p,q)}_z/\epsilon_z$, $\tau^{(p,q)}_x=h^{(p,q)}_x/J_x$. That is to say, $\left(\widetilde{T}^{(p,q)}\right)^\dag$ can be achieved through the time evolution operator
\bea
e^{-i\tau^{(p,q)}_x H^{(q+1,q)}_x}e^{-i\tau^{(p,q)}_zH^{(q+1,q)}_z}. 
\nn
\eea

It is straightforward to show that $U^{(n+1,n)}$ and its hermitian conjugate can also be obtained through time evolution operators, i.e., 
\bea
U^{(n+1,n)} &=& e^{-i\tau^u_z H^{(n+1,n)}_z}e^{-i\tau^u_x H^{(n+1,n)}_x}, \nn \\
\left(U^{(n+1,n)}\right)^\dag &=& e^{-i\tau^u_x H^{(n+1,n)}_x }e^{-i\tau^u_z H^{(n+1,n)}_z }, \nn
\eea
with the evolution time $\tau^u_z=\pi/(2\epsilon^z)$ and $\tau^u_x=-\pi/(2J^x)$.
Therefore, we finally have
\bea
U(m) = U_{\rm diag}\times\left(\widetilde{T}^{(2,1)}\right)^\dag\prod^m_{p=3}\left[\left(\prod^{2}_{q=p-1}\left(U^{(q+1,q)}\right)^\dag\right) \right. \nn \\
 \left. \times  \left(\prod^{p-2}_{q=1}\left(\widetilde{T}^{(p,q)}\right)^\dag U^{(q+2,q+1)}\right)\left(\widetilde{T}^{(p,p-1)}\right)^\dag\right] . \qquad
\eea
We see that in order to build a general $m\times m$ Haar-random unitary $U(m)$, both the number of $H_x$ and $H_z$ gates we need are $(m-1)(3m-4)/2$. And also a gate $U_{\rm diag}$ is needed, which can be achieved through evolving the Hamiltonian $H_d=\frac{i}{8\pi}\log(U_{\rm diag})$ with the time $\tau_d=\hbar8\pi/E_R$. In Fig.~\ref{fig:gates}, we show all the Hamiltonians of typical quantum gates can be precisely achieved with our optimization method, and the absolute errors in the tight-binding model compared to the target one is made smaller than $1E-5$.

\medskip 
\ncsection{Acknowledgment} 

We acknowledge helpful discussion with Peter Zoller, Immanuel Bloch, Markus Greiner, and Yu-Ao Chen. 
This work is supported by National Natural Science Foundation of China under Grants No. 11934002,  11774067, 
National Program on Key Basic Research Project of China under Grant No. 2017YFA0304204, 
and Shanghai Municipal Science and Technology Major Project (Grant No. 2019SHZDZX04). 
Xingze Qiu acknowledges support from National Postdoctoral Program for Innovative Talents of China under Grant No. BX20190083.

 \medskip 
\ncsection{Author Contributions}

{X.L. conceived the main idea in discussion with X.D.Q.; X.Z.Q. and J.Z. developed the methods and performed numerical calculations. All authors contributed in completing the paper.}

\medskip 
\ncsection{Additional Information}

The authors declare no competing interests. Correspondence and requests for material should be sent to xiaopen\_li@fudan.edu.cn. Data is available upon reasonable request.

\bibliographystyle{naturemag}
\bibliography{ref.bib}

\end{document}